\documentclass[aps,preprint,graphicx,unsortedaddress,showpacs]{revtex4}

\renewcommand{\baselinestretch}{1.7}
\usepackage{hyperref}
\usepackage[english]{babel}
\usepackage[T1]{fontenc}
\usepackage[latin1]{inputenc}
\usepackage[dvips]{graphicx}
\usepackage{amsmath}
\usepackage{mathenv}
\usepackage{array}
\usepackage{bm}
\usepackage{fancyhdr}
\usepackage{makeidx}
\usepackage{subfigure}
\usepackage{babel}
\usepackage[v2]{xy}
\usepackage{epsfig}
\usepackage{amsfonts}
\usepackage{amssymb}
\usepackage{color}
\hypersetup{backref=true,pagebackref=true,hyperindex=true,colorlinks=true,breaklinks=true,urlcolor=
blue,linkcolor=blue,bookmarks=true,bookmarksopen=true}

\begin{document}
\renewcommand{\baselinestretch}{1.5}
\title{X-Ray Analysis of Oxygen-induced Perpendicular Magnetic Anisotropy in Pt/Co/AlOx trilayers}
\date{\today}
\author{A. Manchon}
\affiliation{SPINTEC, URA 2512 CEA/CNRS, CEA/Grenoble, 38054 Grenoble Cedex 9, France}
\author{S. Pizzini}
%\affiliation{Institut Néel, CNRS, Grenoble, France}
\author{J. Vogel}
%\affiliation{Institut Néel, CNRS, Grenoble, France}
\author{V. Uhl\^{i}r}
\affiliation{Institut Néel, CNRS/UJF, B.P. 166, 38042 Grenoble Cedex 9, France}
\author{L. Lombard}
%\affiliation{SPINTEC, URA 2512 CEA/CNRS, CEA/Grenoble, 38054 Grenoble Cedex 9, France}
\author{C. Ducruet}
%\affiliation{SPINTEC, URA 2512 CEA/CNRS, CEA/Grenoble, 38054 Grenoble Cedex 9, France}
\author{S. Auffret}
%\affiliation{SPINTEC, URA 2512 CEA/CNRS, CEA/Grenoble, 38054 Grenoble Cedex 9, France}
\author{B. Rodmacq}
%\affiliation{SPINTEC, URA 2512 CEA/CNRS, CEA/Grenoble, 38054 Grenoble Cedex 9, France}
\author{B. Dieny}
\affiliation{SPINTEC, URA 2512 CEA/CNRS, CEA/Grenoble, 38054 Grenoble Cedex 9, France}
\author{M. Hochstrasser}
\affiliation{Laboratory for Solid State Physics, ETH Zürich, 8093 Zürich, Switzerland}
\author{G. Panaccione}
\affiliation{Laboratory TASC, INFM-CNR, Area Science Park, S.S.14, Km 163.5, I-34012, Trieste, Italy}

\begin{abstract}
X-ray spectroscopy measurements have been performed on a series of Pt/Co/AlOx trilayers to investigate the role of Co oxidation in the perpendicular magnetic anisotropy of the Co/AlOx interface. It is observed that high temperature annealing modifies the magnetic properties of the Co layer, inducing an enhancement of the perpendicular magnetic anisotropy. The microscopic structural properties are analyzed via X-ray Absorption Spectroscopy, X-ray Magnetic Circular Dichroism and X-ray Photoelectron Spectroscopy measurements. It is shown that annealing enhances the amount of interfacial oxide, which may be at the origin of a strong perpendicular magnetic anisotropy.
\end{abstract}
\maketitle
\maketitle
\clearpage

The discovery of perpendicular magnetic anisotropy (PMA) in Pd/Co multilayers \cite{carcia85} has opened an exciting field of research questioning the fundamental origins of such PMA and the role of interfacial orbital hybridization \cite{daalderop,nakajima}. 
Although the mechanism of interfacial PMA is complex (see Ref.~\onlinecite{daalderop}), it has been clearly shown that spin-orbit coupling plays a key role in PMA. Daalderop {\it et al.} \cite{daalderop} showed theoretically an oscillating behavior of the anisotropy energy as a function of the valence band filling of Co. Schematically, hybridization of the 3$d$ levels of cobalt with the $d$ levels of Ni, Pd or Pt \cite{nakajima} can modify this filling and induce an important anisotropy in the $d$ orbitals at the interface, thus creating an important PMA in thin X/Co multilayers \cite{weller,bruno} (X=Ni,Pt,Pd).\par
It has recently been shown that strong PMA can be obtained in optimally oxydized Pt/Co/MOx trilayers \cite{monso,rodmacq} (M=Al, Cr, Ta, Mg, Ru), starting from unoxidized films with in-plane magnetization. Recent work of Lacour {\it et al.} \cite{lacour} on similar Pt/Co/Al trilayers showed that Co magnetization in their samples already lies out-of-plane even without oxidation. These results are not in contradiction with those of Ref.~\onlinecite{monso,rodmacq}, but are simply explained by the larger role of the PMA at the Pt/Co interface induced by the presence of a Ta buffer layer and a thicker Pt underlayer. This is a well-known phenomenon as described for instance by Ref.~\onlinecite{huang}.\par

The studies presented in Ref.~\onlinecite{monso,rodmacq} and in the present work can be put in parallel with the theoretical work of Oleinik {\it et al.} \cite{oleinik} and Belaschenko {\it et al.} \cite{belaprb} on Co/AlOx/Co and Co/STO/Co magnetic tunnel junctions (MTJ), which consider the role of interfacial oxygen in the enhancement of the magnetic moment of cobalt. The charge transfer which occurs between Co and O enhances the asymmetry of the valence band of Co, increasing the interfacial spin polarization. This charge transfer reduces the energy of Co $d$ orbitals pointing towards O, creating a splitting between in-plane ($d_{xy}$, $d_{x^2-y^2}$) and out-of-plane ($d_{xz}$, $d_{yz}$, $d_{z^2}$) $d$ orbitals (parameter $\Delta$ in Ref.~\onlinecite{bruno}) increasing the crystalline field effect \cite{bruno}. Consequently, one can expect a strong enhancement of PMA due to the presence of Co-O bondings at the interface. X-ray studies achieved by Telling {\it et al.} \cite{telling2004,telling2006} on Co/AlOx/Co MTJs showed that a maximum of Co magnetic moment, corresponding to a maximum in TMR, can be reached by optimally oxidizing the tunnel barrier. Designing high TMR-MTJ using oxygen-induced PMA \cite{rodmacq} requires a deep understanding of this phenomenon.\par

In this article, we investigate the role of interfacial oxidation in the anisotropy crossover induced by annealing of a Pt/Co/AlOx sandwich. The macroscopic magnetic properties are analyzed in parallel with the microscopic structural properties of the samples obtained via X-ray Photoelectron Spectroscopy (XPS), X-ray Absorption Spectroscopy (XAS) and X-ray Magnetic Circular Dichroïsm (XMCD).\par
Pt(3 nm)/Co(0.6 nm/Al(1.6 nm) trilayers were deposited on a thermally oxidized silicon wafer by conventional dc magnetron sputtering with a base pressure of $5\times10^{-8}$ mbar. The sample was oxidized by using oxygen rf plasma with a partial pressure of $3\times10^{-3}$ mbar and a power of 10 W during 20s. One sample was left as-deposited and a second one was annealed at 350°C for 30 minutes. The samples are highly stable so that no capping layer had to be deposited on top of AlOx.\par

\begin{figure}[h]
	\centering
		\includegraphics[width=9cm]{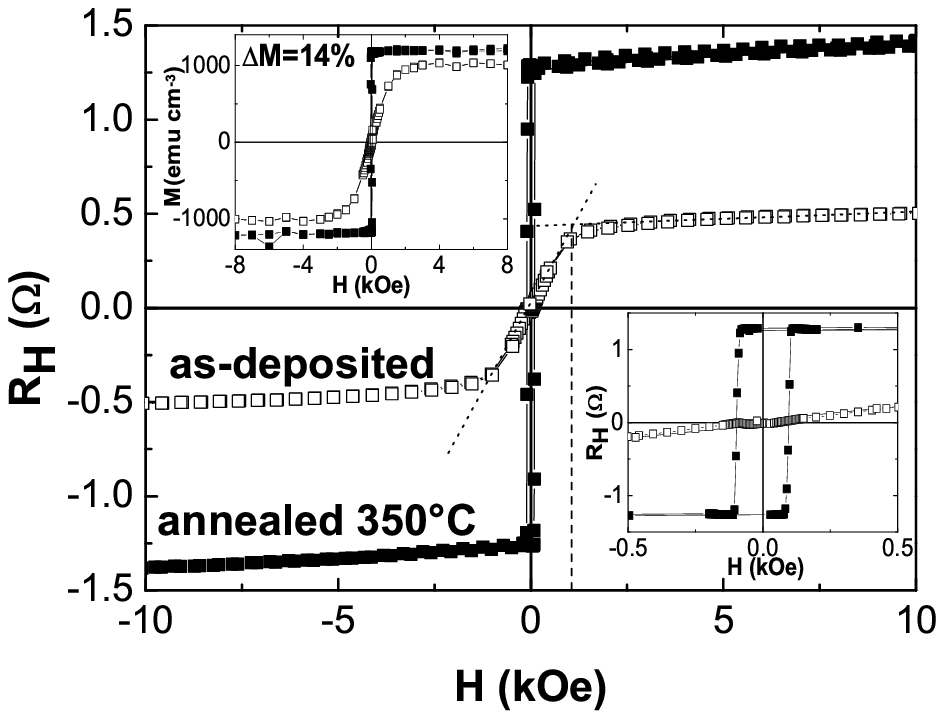}
		\caption{Hall resistance as a function of the applied field for as-deposited (open squares) and annealed (filled squares) samples. The field is applied perpendicular to the plane of the trilayer. Top inset: hysteresis cycles of the samples obtained by SQUID measurements. Bottom inset: zoom of the Hall resistance loops at low field for as-deposited and annealed samples.}
	\label{fig:mag}
\end{figure}

The macroscopic magnetic properties of these samples were studied by extraordinary Hall effect (EHE) in a standard four-probes Hall geometry. The Hall resistance $R_H$ has the form $R_H=R_0H+4\pi R_{EHE}M_z$, where $H$ is the magnetic field (applied perpendicularly to the plane of the sample) and $M_z$ is the out-of-plane component of the sample magnetization. $R_0$ is the ordinary Hall resistance and $R_{EHE}$ is the extraordinary Hall resistance \cite{canedy}. Fig. \ref{fig:mag} shows the hysteresis loops obtained by EHE at room temperature for as-deposited and annealed samples.\par

The EHE measurement of the as-deposited sample gives no hysteresis and zero remanence. Magnetization saturates at high field (|H|>1.4 kOe), where $M_z$ no longer depends on H and the residual slope comes from the ordinary Hall contribution. This demonstrates that the magnetization of the Co layer possesses no effective out-of-plane component at remanence. In fact, EHE measurements with in-plane external field as well as anisotropic magnetoresistance measurements (not shown) demonstrate that the magnetization of the Co layer lies out-of-plane forming a domain structure. Then, one needs to apply a large external field (H=1.4 kOe) to saturate it in the out-of-plane direction. Note that this field is much lower than $4\pi Ms$ ($\approx$ 12 kOe) which is consistent with the existence of PMA at the Co/Pt interface. After annealing, the sample shows a square hysteresis loop with sharp magnetization reversal and a coercivity of 100 Oe (see bottom inset of Fig. \ref{fig:mag}). These features indicate that the perpendicular anisotropy strongly increases in the annealed sample. The perpendicular magnetic anisotropy, estimated from both SQUID and EHE with in-plane external field measurements, is enhanced during the thermal annealing from $H_{an}\approx0.4$ kOe to $H_{an}\approx5.2$ kOe.\par

SQUID measurements applying an out-of-plane external field (top inset of Fig. 1) show that an enhancement of 14\% of the Co magnetic moment is obtained after annealing. Note that the magnetic moment of our ultrathin Co ($\approx$950 emu.cm$^{-3}$) is lower than for bulk Co ($\approx$1400 emu.cm$^{-3}$). The large increase of the EHE magnitude (200\%) in the annealed sample is then mostly attributed to the enhancement of $R_{EHE}$. This point will be adressed elsewhere. \par

To understand the microscopic origin of the anisotropy crossover associated with annealing, we performed soft X-ray Photoelecton Spectroscopy (XPS) of Co 2$p$ levels, as well as X-ray Absorption Spectroscopy (XAS) and X-ray Magnetic Circular Dichroism (XMCD) measurements at the Co $L_{2,3}$ edges. The aim of these element-specific measurements is to correlate the changes in macroscopic magnetic properties induced by annealing with changes in average chemical composition at the Co/Al interface. For the Co thickness used here (0.6 nm), XAS and XMCD give information on the composition and magnetic moments averaged over the Co layer, while XPS will provide information on the electronic structure of Co at the Co/AlOx interface mainly.\par

\begin{figure}[h]
	\centering
		\includegraphics[width=8cm]{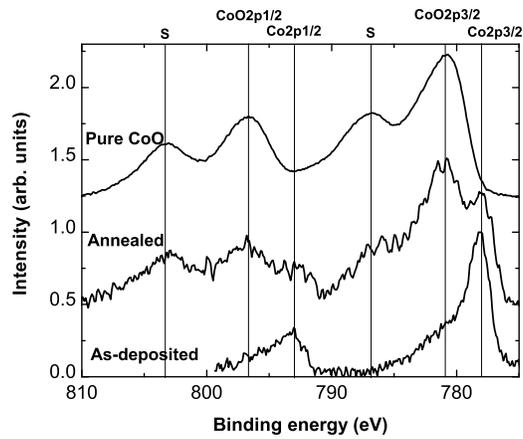}
		\caption{Co 2$p$ XPS spectra for the as-deposited sample (bottom curve), annealed sample (middle curve) and pure CoO (top curve).}
	\label{fig:xps}
\end{figure}

Measurements were carried out at the Advanced Photoelectric-effect Experiments (APE) beamline of the ELETTRA synchroton in Trieste (Italy), using the experimental setup given in the inset of Fig. 3(a). Co 2$p$ XPS measurements were performed setting the incident photon energy at $h\nu=1130$ eV, with an incident angle of $\theta=60$°. Fig. \ref{fig:xps} shows the spectra obtained for as-deposited and annealed samples. The spectrum of pure CoO is given for reference in the top of the same figure. In this spectrum, one can distinguish two main peaks, corresponding to CoO 2$p_{1/2}$ and CoO 2$p_{3/2}$ core levels (lying at 796.3 eV and 781.1 eV respectively) and two satellite peaks (denoted S and lying at 803.3 eV and 786.7 eV resp.) which arise from the charge transfer between O 2$p$ and Co 3$d$. The main peaks are shifted towards larger binding energies with respect to Co metal peaks, due to oxygen environment. The spectrum of the as-deposited sample is similar to that of pure cobalt (see Ref.~\onlinecite{sicot}), with Co 2$p_{1/2}$ and Co 2$p_{3/2}$ lying at the binding energies of 792.5 eV and 778.1eV respectively.\par

The spectrum of the annealed sample clearly shows contributions from both Co and CoO spectra. Gaussian fits (not shown) indicate that the contribution of CoO to the XPS spectra of the annealed sample is larger than 80\%, showing that most of the Co atoms located near the Co/Al interface are bonded to oxygen atoms. Furthermore, the ratio between the estimated areas of the CoO 2$p_{1/2}$ peak (796.3 eV) and its satellite (803.3 eV) gives 0.97, which means that the cobalt oxide is a monoxide \cite{CoO}.\par

\begin{figure}[h]
	\centering
		\includegraphics[width=8cm]{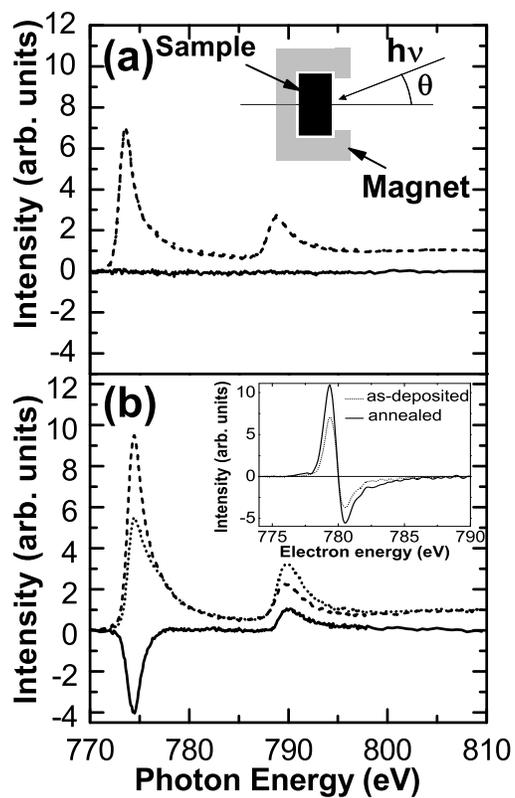}
		\caption{Co $L_{2,3}$ X-ray absorption spectra measured for the as-deposited sample (a) and the annealed sample (b), with no applied magnetic field (H=0) and perpendicular incident beam ($\theta=0$°). The solid curves represent the XMCD signal, obtained from the difference between left (dashed line) and right (dotted line) circularly polarized light spectra. Inset of (a): experimental set-up. Inset of (b): derivative of the XAS spectra of as-deposited (dotted line) and annealed (solid line) samples.} 
	\label{fig:xmcd}
\end{figure}

In order to confirm the previous studies, Co $L_{2,3}$ absorption spectra were obtained by measuring the drain current of the sample via the sample holder. To probe the out-of-plane Co remanent magnetization (no out-of-plane magnetic field could be applied), the sample surface was set perpendicular to the incident X-ray beam (see inset of Fig. \ref{fig:xmcd}(a) with $\theta=0$°).\par

Fig. \ref{fig:xmcd} shows the Co $L_{2,3}$ absorption spectrum measured for the as-deposited and annealed samples, for left (dashed line) and right (dotted line) circular polarization. The spectrum of the as-deposited sample (Fig. 3(a)) shows no dichroism (solid line) in this configuration, indicating that the Co magnetic moment, averaged over the whole Co thickness, has no net component in the direction perpendicular to the surface. This is in agreement with macroscopic measurements. An important dichroism signal is found for the annealed sample, which indicates that it possesses a net, remanent out-of-plane magnetic moment consistent with the EHE measurements. XMCD measurements therefore confirm the increase of the perpendicular magnetic anisotropy associated to annealing.\par

The spectral shape of the Co $L_{2,3}$ absorption spectra is typical of metallic Co (see Ref.~\onlinecite{sicot} for example), both before and after annealing (see the derivatives of the XAS spectra in the inset of Fig. 3(b)). This indicates that the Co chemical environment, averaged over the whole Co layer, does not change significantly upon annealing.  As we have seen, the more surface-sensitive XPS measurements (the collected intensity is exponentially decreasing with a penetration depth of the order of 5-6\AA) indicate, on the other hand, a large oxidation of the Co layer. As a consequence, the  comparison of the two spectroscopic techniques implies that the Co-O environment is specific to the topmost Co/AlOx interface.\par

%No substantial changes in the shape of the XAS spectra are observed upon annealing the sample (see the derivatives of the XAS spectra in the inset of Fig. 3(b)), indicating that the oxygen atoms do not penetrate into the Co layer. An important dichroism signal is found for the annealed sample, which indicates that it posesses a net, remanent out-of-plane magnetic moment consistent with the EHE measurements.\par

%In summary, XAS measurements clearly confirm the increase of perpendicular magnetic anisotropy associated to annealing. On the other hand, these data give the important information that using this technique no significant changes in the Co chemical environment, averaged over the whole Co layer, are observed. Since XPS measurements are more surface sensitive than XAS (the collected intensity is exponentially decreasing with a penetration depth of the order of 5-6\AA), this result indicates that only interfacial Co oxidation appears upon annealing. As a consequence, the comparison of the two spectroscopic techniques implies that the CoO-like environment is specific to the topmost Co/AlOx interface.\par 

We therefore deduce that the enhancement of perpendicular magnetic anisotropy induced by thermal annealing is related to the appearance of a significant density of interfacial Co-O bonds at the Co/AlOx interface. This is the result of temperature-induced diffusion of excess oxygen in the amorphous AlOx structure towards the Co/Al interface.\par

In conclusion, combining bulk and interfacial-sensitive spectroscopy techniques (XAS and XPS), we have shown that interfacial oxidation of Co in Pt/Co/AlOx trilayers is at the origin of strong perpendicular magnetic anisotropy induced by annealing. Thermal annealing favours the oxygen diffusion towards the Co/AlOx interface, without penetration of the O atoms within the Co layer. These studies support the seminal role of oxygen-induced splitting of the Co 3$d$ bands in PMA, as presented at the beginning of this article. The role of the Co/Pt interface have been disregarded in this study because the potential thermally-induced Co-Pt mixing at the Pt/Co interface is expected to reduce the perpendicular magnetic anisotropy, not to enhance it. Ab-initio calculations, aiming at clarifying the role of Co-O hybridization in magnetic anisotropy energy, are presently underway.\par

The authors are grateful for the experimental support and assistance of B. Pang, L. Ranno, O. Fruchart and P. Torelli.

\clearpage

\begin{center}
\textbf{FIGURE CAPTIONS}
\end{center}

\end{document}